\documentclass[aps,pra,twocolumn,letterpaper,superscriptaddress,10pt]{revtex4-2}

\usepackage{amssymb,amsthm,amsmath,amsfonts}
\usepackage{graphicx,mathptmx}
\usepackage[pdftex,dvipsnames,usenames]{xcolor}
\usepackage[colorlinks=true,urlcolor=blue,citecolor=blue,linkcolor=blue]{hyperref}
\usepackage{bbm}
\usepackage[shortlabels]{enumitem}

\usepackage{soul}

\begin{document}

\title{Detector entanglement: Quasidistributions for Bell-state measurements}

\author{Jan Sperling}
	\affiliation{Theoretical Quantum Science, Institute for Photonic Quantum Systems (PhoQS), Paderborn University, Warburger Stra\ss{}e 100, 33098 Paderborn, Germany}
	\email{jan.sperling@upb.de}

\author{Ilaria Gianani}
	\affiliation{Dipartimento di Scienze, Universit\`a degli Studi Roma Tre, Via della Vasca Navale 84, I-00146, Rome, Italy}

\author{Marco Barbieri}
	\affiliation{Dipartimento di Scienze, Universit\`a degli Studi Roma Tre, Via della Vasca Navale 84, I-00146, Rome, Italy}
	\affiliation{Istituto Nazionale di Ottica -- CNR, Largo Enrico Fermi 6, 50125 Florence, Italy}
	
\author{Elizabeth Agudelo}
	\affiliation{Atominstitut, Technische Universit\"at Wien, Stadionallee 2, 1020 Vienna, Austria} 
	\email{elizabeth.agudelo@tuwien.ac.at}

\date{\today}

\begin{abstract}
	Measurements in the quantum domain can exceed classical notions.
	This concerns fundamental questions about the nature of the measurement process itself, as well as applications, such as their function as building blocks of quantum information processing protocols.
	In this paper, we explore the notion of entanglement for detection devices in theory and experiment.
	A method is devised that allows one to determine nonlocal quantum coherence of positive operator-valued measures via negative contributions in a joint distribution that fully describes the measurement apparatus under study.
	This approach is then applied to experimental data for detectors that ideally project onto Bell states.
	In particular, we describe the reconstruction of the aforementioned entanglement quasidistributions from raw data and compare the resulting negativities with those expected from theory.
	Therefore, our method provides a versatile toolbox for analyzing measurements regarding their quantum-correlation features for quantum science and quantum technology. 
\end{abstract}

\maketitle

\section{Introduction}
\label{sec:Introduction}

    Quantum phenomena are understood today as novel resources for advanced quantum operations that constitute the foundation of modern quantum technologies.
    A variety of notions of nonclassicality, such as entanglement, are results of quantum superpositions of states.
    Such quantum interference phenomena, nowadays collectively referred to as quantum coherence, can provide the sought-after resources for quantum information processing \cite{ABC16,SAP17,CG19}.
    While the notion of coherence has a longstanding tradition in quantum optics \cite{M86,G63,TG65,MW95book,VW06book}, only recently have broader concepts of quantum coherence been recognized and extensively studied in the context of operational usefulness in quantum information theory.
    This encompasses entanglement of multipartite quantum states as the essential nonlocal component of quantum coherence \cite{SAP17,CH16}.
    For instance, entanglement is the basis for steering \cite{UCNG20}, as well as generalized notions of conditional quantum correlations \cite{SBDBJDVW16,ASCBZV17}.

    Equally fundamental, yet less frequently addressed, is the matter of the quantumness of measurements.
    Recently, however, this topic has gained considerable momentum, and multiple theoretical methods for the certification of quantum features of detectors were put forward \cite{YDXLS17,BKB19,SL19,SSC19,TR19,BSLKN20,GMSG21}.
    Making the leap from state-based quantum coherence to quantifying the quantum performance of measurement devices is important for measurement-based quantum computation, providing an equivalent approach to state-based information processing \cite{VMDB06,BBDRV09}.
    Beyond its relevance for such application, general observables, determined by so-called positive operator-valued measures (POVMs, that project onto nonclassical states are essential for quantum protocols.
    For example, entangled Bell-state measurements (BSMs) are paramount in quantum teleportation and, by extension, in quantum repeaters for quantum communication via entanglement swapping;
    see, e.g., Ref. \cite{HHGZLHLGGBT22} for a recent experiment.
    In addition, pioneering experiments have reported on the quantumness of measurements \cite{Yokoyama19, Xu2019,Xu2021}.
    For example, experiments have confirmed the noncommutativity of certain observables \cite{PZKB07,ZPKJB09} and proved the incompatibility of quantum measurements with classical statistical models on a quantitative basis.
    Furthermore, fundamental measurement-induced quantum coherence effects of sequential measurements have been investigated \cite{CGSSB19}.
    However, a generally applicable strategy for a theoretical and experimental certification of nonlocal coherence of measurements is still missing.

    In the context of quantum optics, the close relation between entanglement and quantum coherence of multimode light is well known \cite{KSBK02,W02,ACR05}, and quantitative relations between single-mode nonclassicality and multimode entanglement have been established \cite{VS14}.
    In this context, quasiprobabilities are arguably the most essential and widely applied tool for the characterization of quantum states of quantum fields;
    see Ref. \cite{SV20} for a recent review.
    Nonclassical multimode radiation fields are identified through the failure of such quasiprobabilities to find a correspondence in classical probability theory, typically displayed through negativities.
    Even though exceptions exist \cite{DMWS06}, in general, the origin of such negativities, be it single-mode quantum effects or entanglement, cannot be distinguished.
    Moreover, certain notions of quantum coherence are not detectable via quantum-optical quasiprobabilities.
    To mitigate this limitation, a construction of quasiprobabilities for general notions of quantum coherence of states has been formulated \cite{SW18}.
    This includes the theory of entanglement quasiprobabilities, whose negativities are a necessary and sufficient criterion for the identification of entanglement, for either bi- and multipartite states \cite{SW18}.
    Such entanglement quasiprobabilities even found applications in experiments to probe sources of entangled light \cite{SMBBS19}.

    The experience from quantum optics can serve as a guide to further extend detector characterization strategies to modern concepts.
    For example, the nonclassical properties of single-photon detectors have been studied in experiments via quantum-optical quasidistributions \cite{LFCPSREPW09}.
    Such a quasidistribution applies to detectors and is nonnegative for classical detection devices but not necessarily normalized, contrasting quasiprobabilities of states.
    A similar methodology for detector entanglement has not been established or implemented to date.
    Moreover, very recently, the relation between quantum coherence and entanglement of measurements has been studied in theory \cite{KL22}, analogously to the connection of single-mode nonclassicality and entanglement for states.
    Despite this intriguing relationship, however, the approach provides neither a practical nor an intuitive tool for the quantitative assessment of detector entanglement akin to negativities in quasiprobabilities.

    In this paper we introduce and implement a methodology for the entanglement characterization of POVMs in terms of quasidistributions.
    This allows us to assess the entanglement of detection devices on the basis of negativities in those distributions, constituting a necessary and sufficient method to detect entanglement of measurements.
    Using data from detector tomography, we present in great detail the reconstruction of such quasidistributions for general two-qubit measurements in experiments.
    The resulting negativities of this treatment are then compared with the predictions for ideal BSMs to assess the quality of detector entanglement.
    By mixing POVM elements, we further show that non-entangled measurements are accompanied by nonnegative distributions.
    Moreover, a probe-state method is devised as a sufficient criterion to probe POVM entanglement, which is applied theoretically to study qudits and multipartite settings.
    Thereby, we provide a practical toolbox for studying the quantum performance of detectors with respect to their entanglement features for fundamental studies in quantum science and applications in quantum technology.

\section{POVM entanglement}

	In this section we establish the notion of entanglement of detection devices.
	The paper performs a different data analysis of the experiment reported in  Ref. \cite{RGMSSGB17}, implementing a detector tomography for BSMs, and develops the theory of entanglement quasidistributions for detectors, being based on entanglement quasiprobabilities for bi- and multipartite states \cite{SW18}.
	An experimental reconstruction of entanglement quasiprobabilities for a Bell state was carried out \cite{SMBBS19}.
    Also, the role of complex numbers for the notion of entanglement was experimentally explored in this manner by studying two-rebit states \cite{PDBBSS21}.
	Still, to date, experimental entanglement characterization of POVMs carried out using the approach of entanglement quasiprobabilities is lacking.
	A key feature of our approach is that detector entanglement is intuitively displayed via negativities in joint distributions of POVM elements.
    The underlying method for two qubits employs the two-qubit state representation known as standard form \cite{LMO06}, being a correlation-diagonal representation in Pauli matrices, which is discussed later.

	We formulate the formal aspects of POVM entanglement in Sec. \ref{subsec:Preliminaries}.
	Theoretical expectations for BSMs are discussed in Sec. \ref{subsec:Bellmeasurements}.
	The experiment under study is described in Sec. \ref{subsec:experiment}.
    We conclude this section with an outline of the remainder of this work, Sec. \ref{subsec:OutlineRemainder}.

\subsection{Defining POVM entanglement}
\label{subsec:Preliminaries}

	Let $\Pi_k$ be an element of a POVM, obeying $\forall k:\Pi_k\geq0$ and $\sum_k\Pi_k=\mathbbm 1$.
	As it applies to the experiment under consideration, we restrict ourselves to finite-dimensional Hilbert spaces for the sake of simplicity, especially, two qubits in the following sections.
	Similarly to the definition of separable states \cite{W89}, we say that a POVM is separable if the decomposition
	\begin{equation}
	\label{eq:LocDecomp}
		\Pi_k=\sum_{a,b} Q_k(a,b) |a\rangle\langle a|\otimes|b\rangle\langle b|,
	\end{equation}
	in which $Q_k$ is a nonnegative joint distribution, holds true for all $k$.
	If this is not the case, we say the detection is entangled.
	Since local projectors form a generating set of the entire space of operators \cite{STV98}, the above decomposition is always possible when relaxing the nonnegativity constraint.
	Specifically, $\Pi_k$ is entangled if $Q_k\ngeq0$, meaning there exists an entry $Q_k(a,b)<0$ for at least one pair $(a,b)$.
    Here, the word quasidistributions, rather than quasiprobability, is appropriate since a POVM element $\Pi$ is not necessarily normalized, $\mathrm{tr}(\Pi)\neq1$, and quasiprobabilities require a unit normalization.
	In general, the representation via such quasidistributions is not unique \cite{SW18}.
	The construction of optimal entanglement quasiprobabilities that ensure nonnegativity for the separable states was derived in Ref. \cite{SW18}.
	This method straightforwardly extends to POVMs as $\Pi/\mathrm{tr}(\Pi)$ describes a quantum state, being possible in finite-dimensional spaces where $\mathrm{tr}(\Pi)<\infty$ is always obeyed.

    For the scenario of two qubits and using local transformations $T^{(A)}\otimes T^{(B)}$, two-qubit states can be put into the so-called standard form $[T^{(A)}\otimes T^{(B)}]\rho[T^{(A)}\otimes T^{(B)}]^\dag=\sum_w\rho_w\sigma_w^{\otimes 2}$, which is diagonal in the Pauli matrices $\sigma_w$, $w\in\{0,x,y,z\}$ \cite{LMO06}.
	This principle extends to general two-qubit POVM elements $\Pi$.
	How to obtain this standard form from data was derived in Ref. \cite{SMBBS19} and is the basis to obtain optimal quasidistributions via solutions of the so-called separability eigenvalue equations \cite{SW18}.
	Thereby, optimal quasidistributions for entanglement can be computed that are negative if and only if the POVM element is entangled, extending beyond two qubits too \cite{SW18}.

\begin{figure*}
	\includegraphics[width=.35\textwidth]{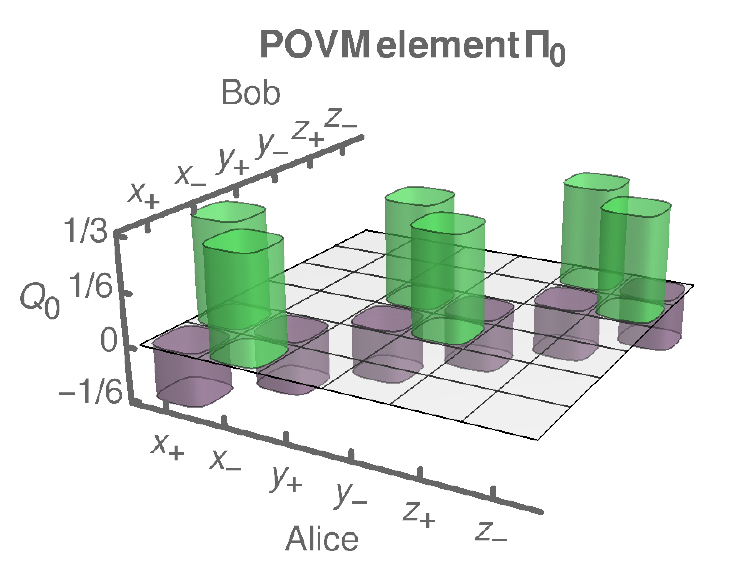}
	\hfill
	\includegraphics[width=.35\textwidth]{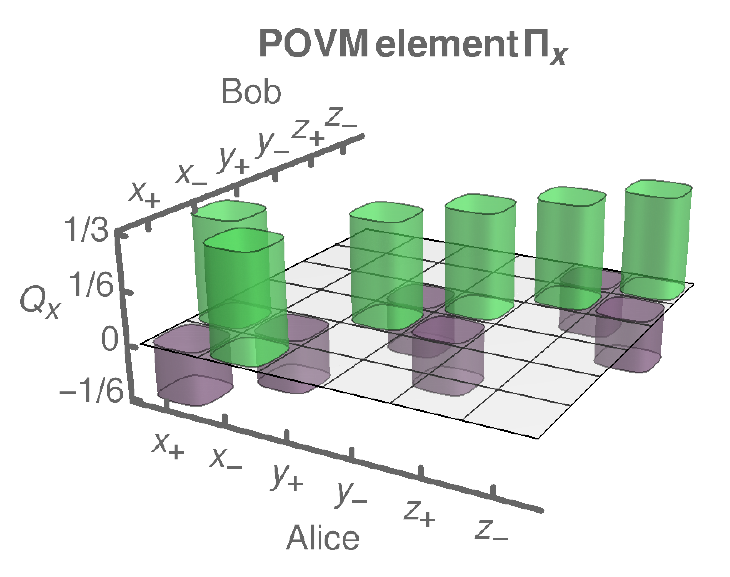}
	\hfill
	\includegraphics[width=.25\textwidth]{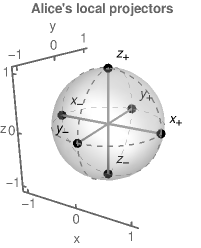}
	\\\vspace{2ex}
	\includegraphics[width=.35\textwidth]{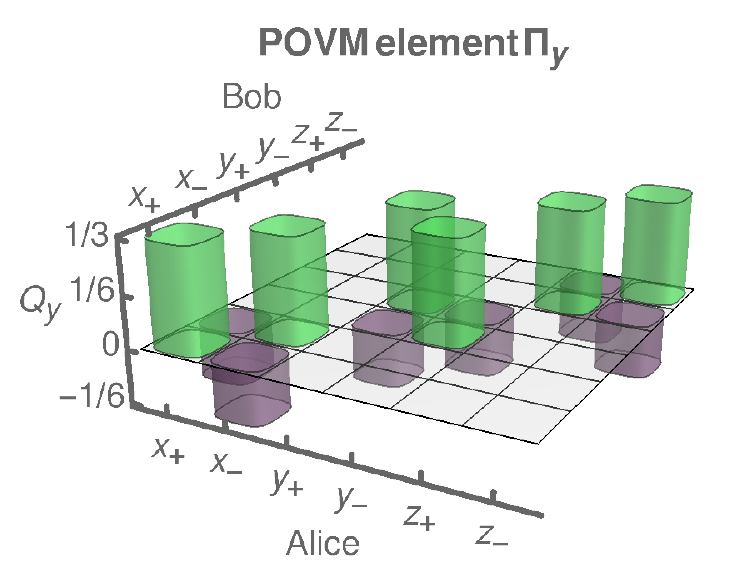}
	\hfill
	\includegraphics[width=.35\textwidth]{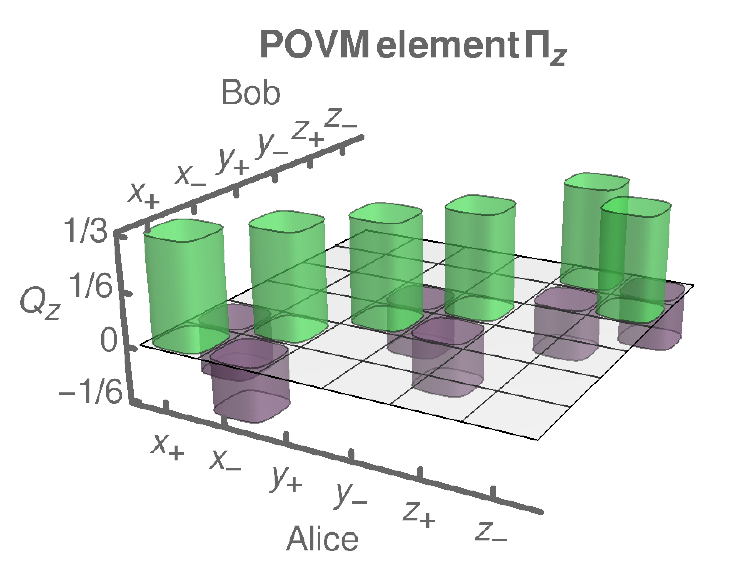}
	\hfill
	\includegraphics[width=.25\textwidth]{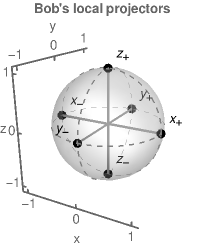}
	\caption{%
		Ideal entanglement quasidistributions $Q_k$ for $k\in\{0,x,y,z\}$ [Eq. \eqref{eq:QuasiStdform}] for Bell-projection POVM elements [Eq. \eqref{eq:BellIdealquasi}].
		Positive contributions ($+1/3$) display classical detector correlations while negativities ($-1/6$) are an unambiguous certification of POVM entanglement.
		Local projectors $|w_\pm\rangle\langle w_\pm|$ (left Bloch-sphere plots) are eigenstates of the Pauli matrices $\sigma_w$ for $w\in\{x,y,z\}$, allowing for a decomposition of the POVM elements according to Eq. \eqref{eq:LocDecomp}.
	}\label{fig:BellIdealquasi}
\end{figure*}

    To clarify, the entanglement of a measurement device is here defined through one or multiple POVM elements that are not nonnegative combinations of local product projectors [Eq. \eqref{eq:LocDecomp}].
    As demonstrated later (Sec. \ref{subsec:witnessing}), this further means that for certain states detection outcomes are possible that are incompatible with local (i.e., separable) POVMs alone.
    This is achieved by so-called entanglement-probing states for nonlocal POVM elements, mirroring the entanglement of states when certified through entanglement witnesses \cite{HHHH09}.
    We emphasize that our approach, despite this similarity, characterizes the entanglement of the measurement scheme under study, as represented through its POVM.
    In addition, in our following experimental study, the possibility to record outcomes that are incompatible with local detection devices is uniquely identified through negativities in the previously discussed entanglement quasidistributions $Q_k$ of POVM elements $\Pi_k$.

\subsection{Predictions for BSMs}
\label{subsec:Bellmeasurements}

	Suppose the standard form $\Pi=\sum_{w\in\{0,x,y,z\}} \pi_w \sigma_w^{\otimes 2}$.
    Then the optimal quasidistribution in a compact form reads \cite{SW18,SMBBS19}
	\begin{equation}
	\label{eq:QuasiStdform}
		Q\left(w^{(A)}_{\pm^{(A)}},w^{(B)}_{\pm^{(B)}}\right)
		=\left(
			\frac{q}{3}+|\pi_{w^{(A)}}|\pm^{(A)}\pm^{(B)}\pi_{w^{(A)}}
		\right)\delta_{w^{(A)},w^{(B)}},
	\end{equation}
	with the parameter $q=\pi_0-|\pi_x|-|\pi_y|-|\pi_z|$ and the Kronecker symbol $\delta$.
	Furthermore, $w_\pm$ labels the eigenstates of Pauli operators, $\sigma_w|w_\pm\rangle=\pm|w_\pm\rangle$, and the superscripts determine Alice's $(A)$ and Bob's $(B)$ subsystems, including signs of their eigenvalues.
	Note that $q\geq0$ and $q<0$ apply to separable and inseparable operators, respectively \cite{SW18}.

	As an example, consider a BSM, represented by the set $\{\Pi_0,\Pi_x,\Pi_y,\Pi_z\}$.
	Each element $\Pi_w$ is a projector $\Pi_w=|\psi_w\rangle\langle\psi_w|$ for a Bell state $|\psi_w\rangle$ that is already in standard form.
	That is, we have
	\begin{eqnarray}
	\label{eq:BellIdealquasi}
	\begin{aligned}
		\Pi_0=&\frac{\sigma_0^{\otimes 2}-\sigma_x^{\otimes 2}-\sigma_y^{\otimes 2}-\sigma_z^{\otimes 2}}{4},
		\\
		\Pi_x=&\frac{\sigma_0^{\otimes 2}-\sigma_x^{\otimes 2}+\sigma_y^{\otimes 2}+\sigma_z^{\otimes 2}}{4},
		\\
		\Pi_y=&\frac{\sigma_0^{\otimes 2}+\sigma_x^{\otimes 2}-\sigma_y^{\otimes 2}+\sigma_z^{\otimes 2}}{4},
		\\\text{and}\quad
		\Pi_z=&\frac{\sigma_0^{\otimes 2}+\sigma_x^{\otimes 2}+\sigma_y^{\otimes 2}-\sigma_z^{\otimes 2}}{4}
	\end{aligned}
	\end{eqnarray}
	for the Bell states $|\psi_0\rangle=(|0\rangle\otimes|1\rangle-|1\rangle\otimes|0\rangle)/\sqrt 2$, $|\psi_x\rangle=(|0\rangle\otimes|0\rangle-|1\rangle\otimes|1\rangle)/\sqrt 2$, $|\psi_y\rangle=(|0\rangle\otimes|0\rangle+|1\rangle\otimes|1\rangle)/\sqrt 2$, and $|\psi_z\rangle=(|0\rangle\otimes|1\rangle+|1\rangle\otimes|0\rangle)/\sqrt 2$, respectively, using the computational basis $\{|0\rangle,|1\rangle\}$.
	Applying Eq. \eqref{eq:QuasiStdform}, the resulting ideal quasidistributions for BSMs are depicted in Fig. \ref{fig:BellIdealquasi}.
	The negativities in those quasidistributions tell us that projective measurements onto Bell states are in fact entangled.
	Those entangled POVM elements of ideal Bell projectors will be compared with our experimental reconstruction later in this paper.

\subsection{Experimental setup}
\label{subsec:experiment}

\begin{figure}
    \includegraphics[width=.95\columnwidth]{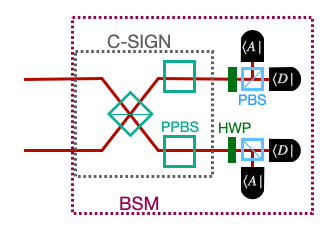}
    \caption{%
        Sketch of the detection scheme that implements a BSM.
        A comprehensive characterization of our realization can be found in Ref. \cite{RGMSSGB17}.
        Because of the final detection in the $D$-$A$ basis, the POVM elements are labeled as $\Pi_{AA}$, $\Pi_{AD}$, $\Pi_{DA}$, and $\Pi_{DD}$ throughout this work, indicating between which detectors coincidences have been recorded.
    }\label{fig:experiment}
\end{figure}

    Consider the entangling detector (Fig. \ref{fig:experiment}), which is based on the use of a photonic control-sign gate (C-SIGN) for polarization qubits \cite{LWPRGOPW05,SKWUZW09,RGMSSGB17,GTCJLBS20}.
    The C-SIGN gate acts on a pair of computational two-qubit basis states as $|j_A\rangle\otimes|j_B\rangle\mapsto(-1)^{j_A\cdot j_B}|j_A\rangle\otimes|j_B\rangle$ for $j_A,j_B\in\{0,1\}$, introducing a $\pi$-phase shift when both qubits take the value true $j_A=j_B=1$.
    Two photons from degenerate spontaneous down-conversion arrive at a partially polarizing beam splitter (PPBS), whose transmittivities are $T_H = 1$ for the horizontal ($H$) component and $T_V = 1/3$ for the vertical ($V$) component;
    therefore, quantum interference can only occur for vertical components, resulting in the desired state-dependent phase shift.
    Two extra beam splitters of this kind, rotated by $90^\circ$, are inserted into the two output ports in order to balance polarization-dependent loss \cite{PGHWP17}.
    The gate works in postselection, accepting only events leading to a coincidence between the two outputs.

    The action of the gate leads the entangled states $\sqrt{2}^{-1/2}(\vert z_+x_+ \rangle \pm \vert z_-x_- \rangle)$ to $\vert x_\pm x_+ \rangle$ and similarly $\sqrt{2}^{-1/2}(\vert z_+x_- \rangle \pm \vert z_-x_+ \rangle)$ to $\vert x_\pm x_- \rangle$. 
    Considering the polarization encoding in Table \ref{tab:LocBases}, this implies that the four states in the Bell basis can be discriminated after the gate by a separable measurement in the diagonal basis.
    In combination, we expect POVM elements to correspond to projectors on Bell states \cite{RGMSSGB17}.
    The main factors causing a departure from the ideal can be identified in the actual values of $T_H$ and $T_V$, imperfect visibility, and local phase shifts.
    These cause not only mixtures but also an unbalance of the expected probabilities.
    A complete characterization of the BSM experiment was presented in Ref. \cite{RGMSSGB17};
    we use the same set of data for our analysis of quasidistributions here. 
    
\begin{table}
	\caption{%
		Computational bases (first column) as chosen for Alice (second column) and Bob (third column) in terms of polarization states.
		Also, the relation to eigenstates $|w_\pm\rangle$ of Pauli matrices $\sigma_w$ are provided.
		(Note that irrelevant global phases are not included.)
	}\label{tab:LocBases}
	\begin{tabular}{ccc}
		\hspace*{.3\columnwidth}&\hspace*{.3\columnwidth}&\hspace*{.3\columnwidth}
		\\[-2ex]
		\hline\hline
		Computational & Alice & Bob
		\\
		\hline
		$|0\rangle=|z_+\rangle$ & $|H\rangle$ & $|D\rangle$
		\\
		$|1\rangle=|z_-\rangle$ & $|V\rangle$ & $|A\rangle$
		\\
		$\frac{|0\rangle+|1\rangle}{\sqrt2}= |x_+\rangle$ & $|D\rangle$ & $|H\rangle$
		\\
		$\frac{|0\rangle-|1\rangle}{\sqrt2}=|x_-\rangle$ & $|A\rangle$ & $|V\rangle$
		\\
		$\frac{|0\rangle+i|1\rangle}{\sqrt2}=|y_+\rangle$ & $|L\rangle$ & $|R\rangle$
		\\
		$\frac{|0\rangle-i|1\rangle}{\sqrt2}=|y_-\rangle$ & $|R\rangle$ & $|L\rangle$
		\\
		\hline\hline
	\end{tabular}
\end{table}

\subsection{Preliminary summary and outline}
\label{subsec:OutlineRemainder}

    Analogous to the notion of inseparability of states \cite{W89}, the notion of inseparable POVMs was established in this section, which naturally extends to more than two parties.
    Since even nonlocal operators can be expanded via products of local operators \cite{STV98}, we argued that entangled POVM elements may be expressed in this manner, however, requiring negative expansion coefficients which are not needed for separable detectors.
    This defines the concept of entanglement quasidistributions of POVMs,
    and negativities in such distributions are a necessary and sufficient criterion for entanglement of detection devices.
    This approach also unifies quasiprobabilities for entangled states \cite{SW18} with quasidistributions for detector entanglement.
    As examples with specific relevance for the continuation of this work, we considered BSMs in two-qubit systems.
    An additional probe-state method to witness POVM entanglement in multipartite qudit systems is introduced at the end of this work, complementing the quasidistribution approach.

    In the remainder of this paper, the specific steps from raw data to entanglement quasidistributions for a BSM are laid out, further applying to arbitrary two-qubit measurements.
    Specifically, the data processing for the detector tomography is described in Sec. \ref{sec:DataProcessingI}.
	Further processing then yields the sought-after entanglement quasidistributions in Sec. \ref{sec:DataProcessingII}.
	A concluding discussion is given in Sec. \ref{sec:Conclusion}, which includes a comparison with our theoretical predictions.
	For instance, the quality of the BSM is assessed by analyzing the maximal negativities, the negativities with the highest statistical significance, and the total negativities.
	Also, a comparison with separable detectors is carried out, where non-entangled POVMs can be straightforwardly mimicked by mixing data to eliminate quantum correlations.
	This leads to a description of POVMs via quasiprobabilities with negativities in joint distributions as unique signatures of its entanglement.


\section{Data processing I: Detector tomography}
\label{sec:DataProcessingI}

	In this section, we start with presenting the measured data in Sec. \ref{subsec:RawData},
	suitable local computational bases are established in Sec. \ref{subsec:CompBases},
	the POVM is reconstructed in Sec. \ref{subsec:CorrPOVM},
	and corrections for unphysical features of POVM elements are discussed in Sec. \ref{subsec:IndefCorrection}.

\subsection{Raw data}
\label{subsec:RawData}

\begin{figure}
	\includegraphics[width=.45\columnwidth]{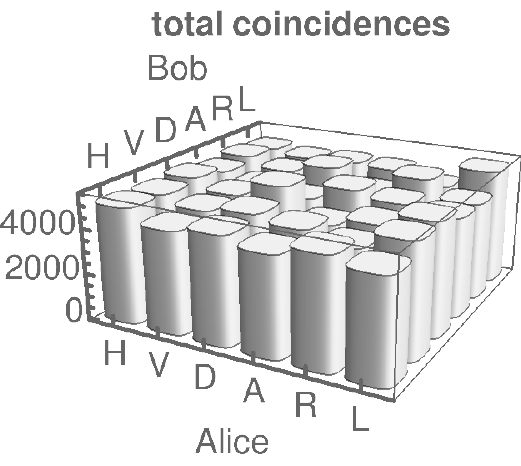}
	\\[2ex]
	\includegraphics[width=.45\columnwidth]{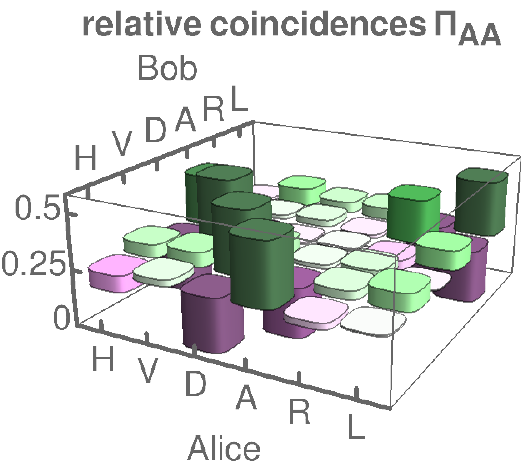}
	\quad
	\includegraphics[width=.45\columnwidth]{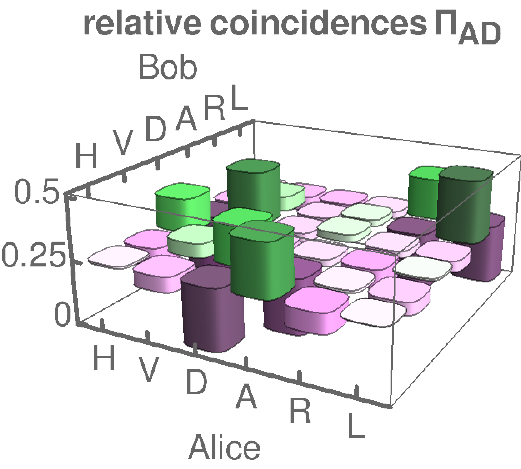}
	\\[2ex]
	\hfill
	\includegraphics[width=.45\columnwidth]{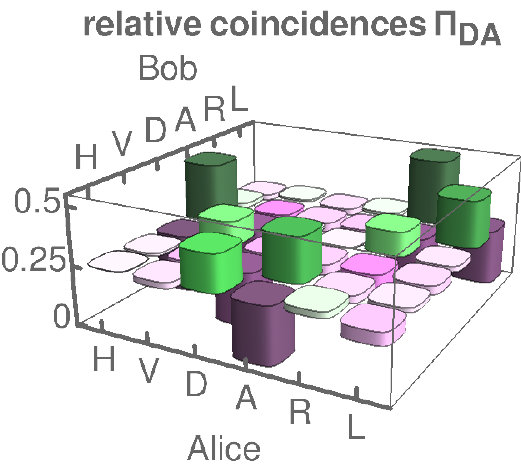}
	\quad
	\includegraphics[width=.45\columnwidth]{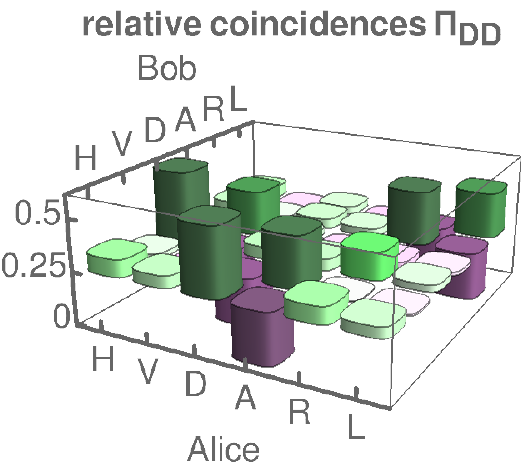}
	\caption{%
		Raw data in the form of the total number of measured coincidences from both detectors (top plot) and the resulting relative coincidences for each POVM element $\Pi_k$.
		Axes label the input product states for horizontal ($H$), vertical ($V$), diagonal ($D$), antidiagonal ($A$), right-circular ($R$) and left-circular ($L$) polarization for realizing the detector tomography.
		Note that bars for relative coincidences are filled to the value $1/4$ to easily distinguish between above- and below-average count rates, i.e., the deviation from uniformity of the total counts distributed among the four individual POVM elements.
	}\label{fig:RawData}
\end{figure}

	Data for each POVM element are recorded for product states $|a\rangle\otimes|b\rangle$ with known polarization to implement a detector tomography.
	For each element, the coincidence counts $E_k(a,b)$ for $k\in\{AA,AD,DA,DD\}$ can be summed to obtain the total counts for each probe state, $E(a,b)=\sum_k E_k(a,b)$.
	Thereby, relative frequencies
	\begin{equation}
		\label{eq:RelFreq}
		p_k(a,b)=\frac{E_k(a,b)}{E(a,b)}
		=\mathrm{tr}\left(\Pi_k|a\rangle\langle a|\otimes|b\rangle\langle b|\right)
	\end{equation}
	can be defined, yielding the probabilities for the following reconstruction.
	Those total counts and relative coincidences are shown in Fig. \ref{fig:RawData}, representing our raw data.
	The probe states for both subsystems comprise
		$|H\rangle$,
		$|V\rangle$,
		$|D\rangle=(|H\rangle+|V\rangle)/\sqrt{2}$,
		$|A\rangle=(|H\rangle-|V\rangle)/\sqrt{2}$,
		$|R\rangle=(|H\rangle-i|V\rangle)/\sqrt{2}$,
		and
		$|L\rangle=(|H\rangle+i|V\rangle)/\sqrt{2}$,
	being the common mutually unbiased bases of a polarization qubit.

\subsection{Local computational bases}
\label{subsec:CompBases}

	As outlined in Ref. \cite{RGMSSGB17}, it is convenient to consider well-chosen local bases.
	That is, Alice uses a horizontal-vertical basis and Bob employs a diagonal-antidiagonal one.
	This choice has no effect on the entanglement but changes representations of Pauli matrices that are used to determine correlations and that are formulated in terms of the computational basis $\{|0\rangle,|1\rangle\}$:
	    $\sigma_z=|0\rangle\langle 0|-|1\rangle\langle 1|$,
	    $\sigma_x=|0\rangle\langle 1|+|1\rangle\langle 0|$,
	    and $\sigma_y=i|1\rangle\langle0|-i|0\rangle\langle 1|$.
	In Table \ref{tab:LocBases}, the choices of bases for the measured data are provided.
	With that, local Pauli matrices can be straightforwardly obtained,
	\begin{equation}
	    \label{eq:PauliExpansion}
        \sigma_w=|w_+\rangle\langle w_+|-|w_-\rangle\langle w_-|
		\quad\text{for}\quad
		w\in\{x,y,z\},
    \end{equation}
	which is relevant for determining the correlations $\mathrm{tr}(\Pi_k \sigma_{w^{(A)}}\otimes\sigma_{w^{(B)}})$ for the POVM elements $\Pi_k$ under study.
	For completeness, the $2\times2$ identity can be expressed symmetrically as
	\begin{equation}
	\label{eq:IdentityExpansion}
	\begin{aligned}
		\sigma_0=&\frac{1}{3}\sigma_0+\frac{1}{3}\sigma_0+\frac{1}{3}\sigma_0
		\\
		=&\frac{1}{3}\sum_{w\in\{x,y,z\}}\left(
			|w_+\rangle\langle w_+|+|w_-\rangle\langle w_-|
		\right).
	\end{aligned}
	\end{equation}

\subsection{Correlation matrix and POVM reconstruction}
\label{subsec:CorrPOVM}

	Our goal is now to decompose the elements $\Pi_k$ in terms of Pauli matrices, $\Pi_k=\sum_{w^{(A)},w^{(B)}\in\{0,x,y,z\}}\pi_{w^{(A)},w^{(B)}|k}\sigma_{w^{(A)}}\otimes\sigma_{w^{(B)}}$, also defining a Pauli-correlation matrix $C_k=[\pi_{w^{(A)},w^{(B)}|k}]_{w^{(A)},w^{(B)}\in\{0,x,y,z\}}$.
	This further yields the computational basis expansion of $\Pi_k$ from the informationally complete set of measurements.
	For a measured probe state, e.g., $|a\rangle\otimes|b\rangle=|w^{(A)}_{\pm^{(A)}}\rangle\otimes|w^{(B)}_{\pm^{(B)}}\rangle$, we can use Eq. \eqref{eq:RelFreq} and the bases  in Table \ref{tab:LocBases} to obtain the desired coefficients from the data in Fig. \ref{fig:RawData}.
	To this end, we can define the matrix of relative coincidences $P_k=[p_k(a,b)]_{a,b\in\{H, V, D, A, R, L\}}$ and sampling matrices
	\begin{equation}
	\label{eq:SampMat}
	\begin{aligned}
		S^{(A)}=&\begin{bmatrix}
			\frac{1}{3} & \frac{1}{3} & \frac{1}{3} & \frac{1}{3} & \frac{1}{3} & \frac{1}{3}
			\\
			0 & 0 & 1 & -1 & 0 & 0
			\\
			0 & 0 & 0 & 0 & -1 & 1
			\\
			1 & -1 & 0 & 0 & 0 & 0
			\\
		\end{bmatrix}
		\\\text{and}\quad
		S^{(B)}=&\begin{bmatrix}
			\frac{1}{3} & \frac{1}{3} & \frac{1}{3} & \frac{1}{3} & \frac{1}{3} & \frac{1}{3}
			\\
			1 & -1 & 0 & 0 & 0 & 0
			\\
			0 & 0 & 0 & 0 & 1 & -1
			\\
			0 & 0 & 1 & -1 & 0 & 0
			\\
		\end{bmatrix}.
	\end{aligned}
	\end{equation}
	Together, those matrices deliver the sought-after expansion coefficients via
	\begin{equation}
	\label{eq:CorrMat}
	\begin{aligned}
		C_k=&\frac{1}{4}S^{(A)}\,P_k\,S^{(B)\mathrm{T}}
		\\
		=&\left[
			\mathrm{tr}\left(
				\Pi_k\tfrac{1}{2}\sigma_{w^{(A)}}\otimes\tfrac{1}{2}\sigma_{w^{(B)}}
			\right)
		\right]_{w^{(A)},w^{(B)}\in\{0,x,y,z\}}
		\\
		=&\left[
			\pi_{w^{(A)},w^{(B)}|k}
		\right]_{w^{(A)},w^{(B)}\in\{0,x,y,z\}},
	\end{aligned}
	\end{equation}
	using orthogonality in the form $\mathrm{tr}(\sigma_w\sigma_{w'})=2\delta_{w,w'}$.
	The coefficients of the sampling matrices in Eq. \eqref{eq:SampMat} describe the relations in Eqs. \eqref{eq:PauliExpansion} and \eqref{eq:IdentityExpansion} as well as Table \ref{tab:LocBases}.

\begin{figure}
	\includegraphics[width=.49\columnwidth]{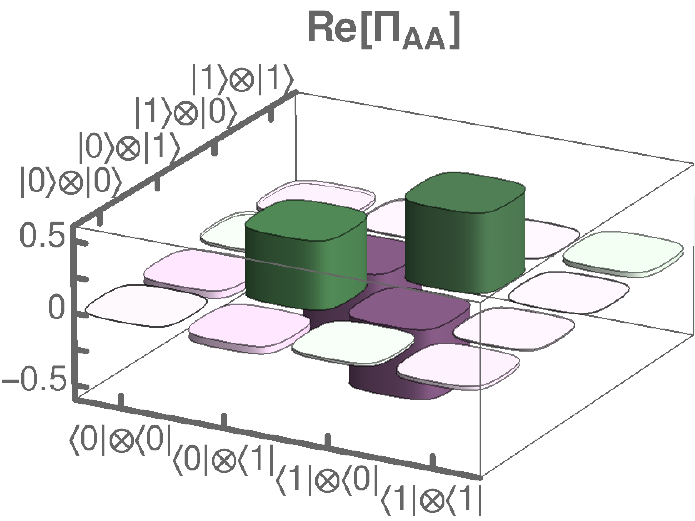}
	\includegraphics[width=.49\columnwidth]{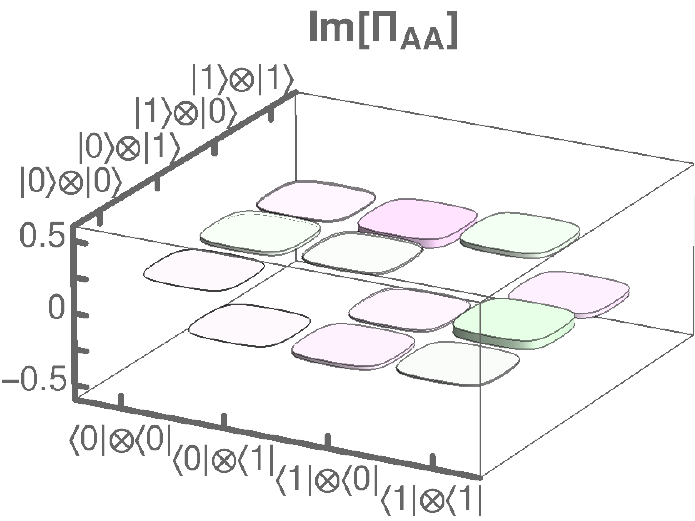}
	\\\vspace{2ex}
	\includegraphics[width=.49\columnwidth]{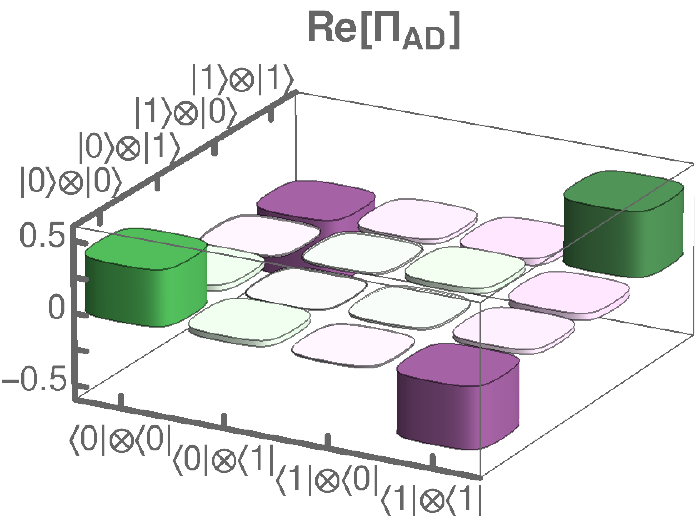}
	\includegraphics[width=.49\columnwidth]{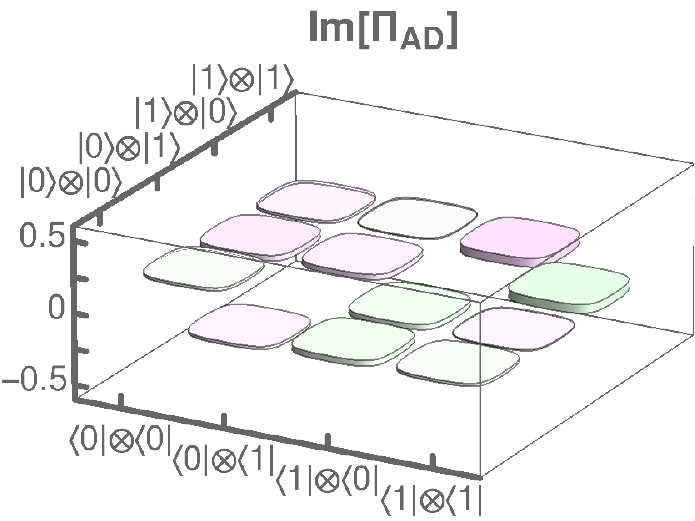}
	\\\vspace{2ex}
	\includegraphics[width=.49\columnwidth]{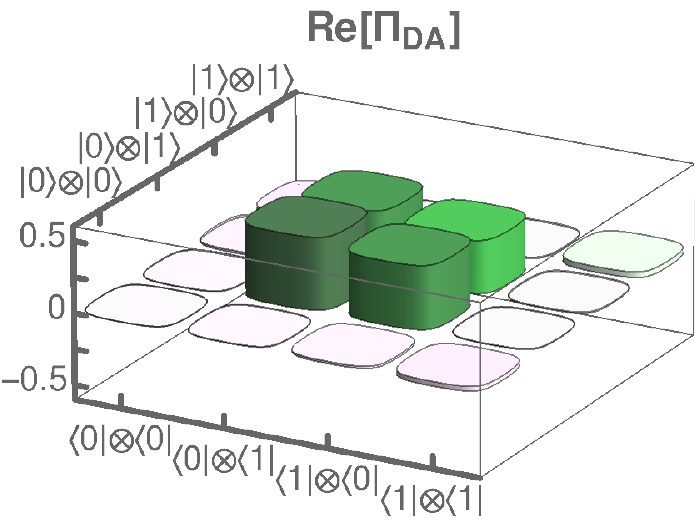}
	\includegraphics[width=.49\columnwidth]{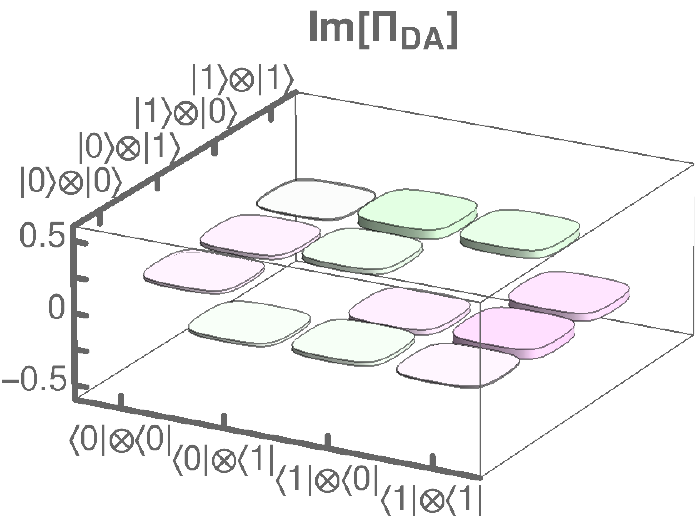}
	\\\vspace{2ex}
	\includegraphics[width=.49\columnwidth]{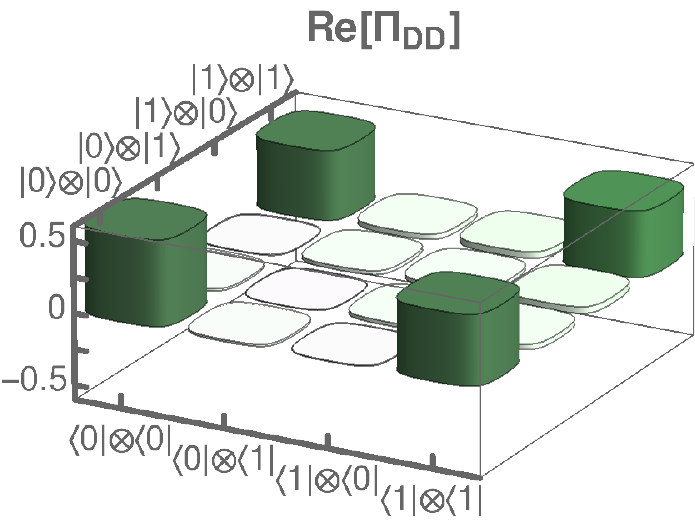}
	\includegraphics[width=.49\columnwidth]{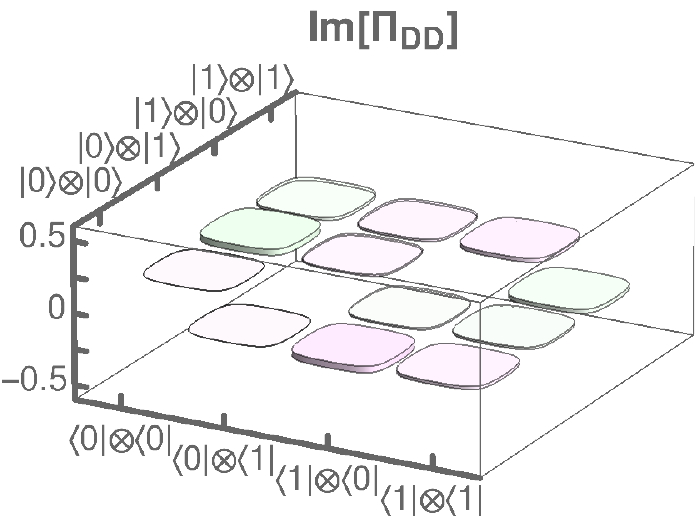}
	\caption{%
		Reconstructed $\Pi_k$ in terms of the real (left column) and imaginary (right column) parts in the computational basis $|k^{(A)}\rangle\langle l^{(A)}|\otimes |k^{(B)}\rangle\langle l^{(B)}|$ for $k^{(A)},l^{(A)},k^{(B)},l^{(B)}\in\{0,1\}$.
		Imaginary contributions are comparably small.
		Real parts show relations to measurements in terms of Bell-state projectors.
		No corrections for imperfections have been carried out to determine the decomposition shown.
	}\label{fig:BasisReconstrct}
\end{figure}

	From the computed coefficients, i.e., elements of $C_k$, one also obtains the basis expansion of $\Pi_k$ as the expansion of the Pauli matrices is known.
	In Fig. \ref{fig:BasisReconstrct}, this bipartite basis expansion as obtained from the data is depicted.
	We find that, from top to bottom, the POVM elements resemble projective measurements for the Bell states $|\psi_0\rangle$, $|\psi_x\rangle$, $|\psi_z\rangle$, and $|\psi_y\rangle$.
	The fidelities with those projectors were reported previously \cite{RGMSSGB17}, using a different reconstruction approach, and are all above $90\%$.

\subsection{Noise addition for indefiniteness}
\label{subsec:IndefCorrection}

	Within the numerical precision ($10^{-9}$), the reconstructed POVM elements satisfy $\sum_k\Pi_k=\mathbbm1$.
	However, the positive semidefiniteness, $\Pi_k\geq0$, is slightly violated, constituting a common issue in tomographic reconstruction schemes.
	This indefiniteness can be easily accounted for without falsely increasing POVM entanglement properties;
	this is discussed in the following.

	For the aforementioned correction, we consider a uniform white-noise addition
	\begin{equation}
		\label{eq:CorrectingSepNoise}
		\Pi_k\mapsto (1-p)\Pi_k+p\frac{1}{4}\mathbbm1
	\end{equation}
	for all four POVM elements and $0\leq p\leq 1$.
	Since the two-qubit identity $\mathbbm1=\sigma_0^{\otimes 2}$ is a product---hence, uncorrelated---operator, the above mixing operation with the separable $\mathbbm1$ cannot increase inseparability.
	Moreover, $\sum_k\Pi_k=\mathbbm1$ is also not influenced by this mapping.
	In the following, the mixing probability $p$ is chosen such that $\Pi_k\geq0$ is simultaneously satisfied for all $k$.
	Importantly, this procedure makes sure that negativities we observe in quasidistributions are a result of entanglement and not a result from slightly unphysical POVM reconstructions.

	Let $-\lambda_\text{max. neg.}$ be the minimal eigenvalue of all POVM elements.
	(We set $\lambda_\text{max. neg.}=0$ if all elements are already positive semidefinite.)
	To further enhance numerical stability, we can add a small extra margin, $\lambda_\text{max. neg.}\mapsto\lambda_\text{max. neg.}+10^{-5}$, implying positive definiteness $\Pi_k>0$.
	In our case, we get $\lambda_\text{max. neg.}\approx 0.05$ in this manner, which is comparably small considering maximal positive eigenvalues that are close to unity.
	Finally, the map in Eq. \eqref{eq:CorrectingSepNoise} results in a proper (i.e., physical) POVM for $p=\lambda_\text{max. neg.}/(\lambda_\text{max. neg.}+1/4)$.
	The thereby obtained POVM is used for further entanglement characterization, despite resulting in reduced quantum correlations because of the extra uncorrelated noise.


\section{Data processing II: Quasidistribution reconstruction}
\label{sec:DataProcessingII}

	In this section, the reconstruction of entanglement quasidistributions is carried out, which is based on the results of the preceding section.
	The transformation of correlation matrices to the standard form is presented in Sec. \ref{subsec:TrafoStdFormI},
	and the thereby implied transformation of local bases states is given in Sec. \ref{subsec:TrafoStdFormII}.
	The propagation of uncertainties via a common Monte Carlo approach is explained for completeness in Sec. \ref{subsec:Uncertainties}.

\subsection{Transformation to standard form}
\label{subsec:TrafoStdFormI}

	As developed in Ref. \cite{SMBBS19}, the numerical transformation of the correlation matrices $C_k$, containing the expansion coefficients of $\Pi_k$ in a Pauli-operator expansion, is a two-step process.
	The first transformation removes local elements such that coefficients for $\sigma_w\otimes\sigma_0$ and $\sigma_0\otimes\sigma_w$ vanish for all $w\in\{x,y,z\}$.
	The second step is a rotation for concluding the diagonalization, meaning that coefficients for $\sigma_{w}\otimes\sigma_{w'}$ become zero for $w\neq w'$.

\begin{figure*}
	\includegraphics[width=.25\textwidth]{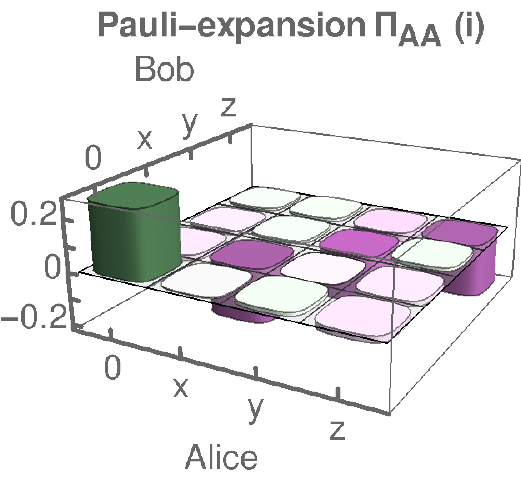}
	\hfill
	\includegraphics[width=.25\textwidth]{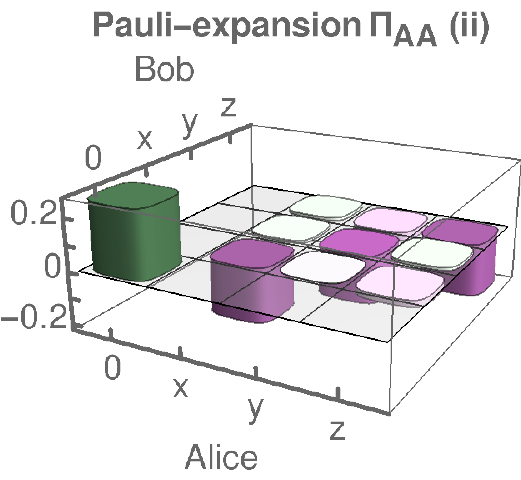}
	\hfill
	\includegraphics[width=.25\textwidth]{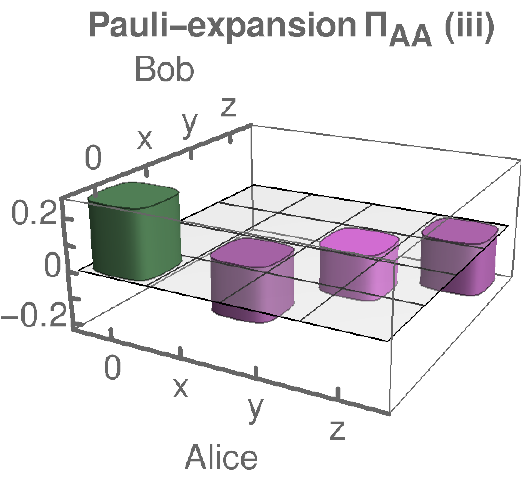}
	\caption{%
		Transformation to the standard form of the correlation matrix $C_{AA}=[\pi_{w,w'|AA}]_{w,w'\in\{0,x,y,z\}}$ that describes the POVM element $\Pi_{AA}$ as an example.
		The initial matrix (i) is transformed such that $\pi_{w,0|AA}=\pi_{0,w'|AA}=0$ holds true for (ii), thus removing local correlations that are given via an identity in one subsystem.
		Then the fully diagonal form in (iii) is obtained by local rotations, resulting in coefficients $\pi_{w,w'|AA}=0$ for $w\neq w'$.
	}\label{fig:CorrelatDiag}
\end{figure*}

\begin{figure*}
	\includegraphics[width=.225\textwidth]{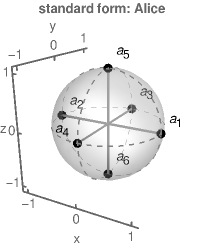}
	\hfill
	\includegraphics[width=.225\textwidth]{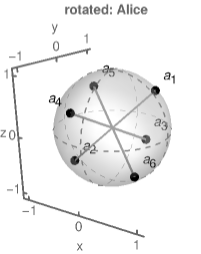}
	\hfill
	\includegraphics[width=.225\textwidth]{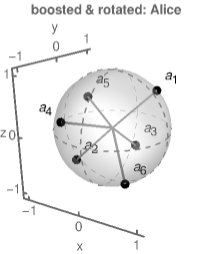}
	\\\vspace{2ex}
	\includegraphics[width=.225\textwidth]{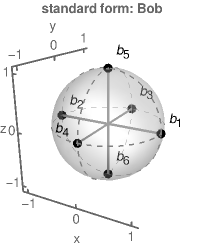}
	\hfill
	\includegraphics[width=.225\textwidth]{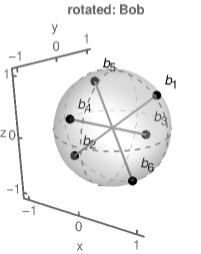}
	\hfill
	\includegraphics[width=.225\textwidth]{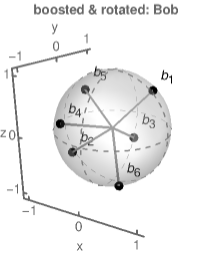}
	\caption{%
		Local state transformation on the Bloch sphere as obtained from the diagonalization steps in Fig. \ref{fig:CorrelatDiag} for $\Pi_{AA}$, but in inverse order from left to right.
		Also compare Eqs. \eqref{eq:ToStandardForm} and \eqref{eq:FromStandardForm} in this context.
		The first transformation constitutes a rotation and the second one is an invertible but not orthogonality-preserving Lorentz boost.
	}\label{fig:LocalStateTransform}
\end{figure*}

	In Fig. \ref{fig:CorrelatDiag}, the two steps are depicted, exemplified for $\Pi_{k}$ with $k=AA$.
	One can see how $C_k$ is successively becoming more diagonal.
	The first step, (i)$\mapsto$(ii), acts like a Lorentz boost operation on the Pauli-expansion \cite{LMO06,SMBBS19}.
	In terms of the operators themselves, this describes a local invertible operation
	\begin{equation}
		\Pi\mapsto \left[L^{(A)}\otimes L^{(B)}\right]\Pi \left[L^{(A)}\otimes L^{(B)}\right]^\dag,
	\end{equation}
	where the inverse for both $L^{(A)}$ and $L^{(B)}$ exists but is generally not unitary.
	The second transformation (ii)$\mapsto$(iii) is a $\mathrm{SO}(3)$ rotation in the Pauli representation.
	It is worth mentioning that the rotations are chosen such that the ordering of magnitudes and signs along the diagonal are preserved, which helps to minimize rotations on the Bloch sphere and preserves directionality to some extent when compared with our initial local basis choice.
	The obtained rotations act as a local unitary on the operators
	\begin{equation}
		\Pi\mapsto \left[U^{(A)}\otimes U^{(B)}\right]\Pi \left[U^{(A)}\otimes U^{(B)}\right]^\dag.
	\end{equation}
	Eventually, we obtain the sought-after standard form to which we can apply the quasidistribution as expressed in Eq. \eqref{eq:QuasiStdform}, likewise
	\begin{equation}
	\label{eq:ToStandardForm}
	\begin{aligned}
		&\left[U^{(A)}L^{(A)}\otimes U^{(B)}L^{(B)}\right]
		\Pi
		\left[U^{(A)}L^{(A)}\otimes U^{(B)}L^{(B)}\right]^\dag
		\\
		=&\sum_{w\in\{0,x,y,z\}}\pi_w\sigma_w^{\otimes 2}
		=\sum_{k,l=1}^6 Q(a_k,b_l)|a_k\rangle\langle a_k|\otimes|b_l\rangle\langle b_l|,
	\end{aligned}
	\end{equation}
	where we relabel the states $[x_+,x_-,y_+,y_-,z_+,z_-]$ as $[a_1,\ldots,a_6]$ and $[b_1,\ldots,b_6]$ for convenience and $Q$ is the solution in standard form \eqref{eq:QuasiStdform}.
	The numerical specifics for determining the boost-like and rotation operations can be found in the Supplemental Material of Ref. \cite{SMBBS19}.

\subsection{Local basis transformations}
\label{subsec:TrafoStdFormII}

	Conversely to the previous relation, we can express the POVM element as through the inverse transformation.
	That is, we have
	\begin{equation}
	\label{eq:FromStandardForm}
	\begin{aligned}
		\Pi=&\sum_{k,l=1}^6 Q(\tilde a_k,\tilde b_l)|\tilde a_k\rangle\langle \tilde a_k|\otimes|\tilde b_l\rangle\langle \tilde b_l|
		\\
		=&\left[U^{(A)}L^{(A)}\otimes U^{(B)}L^{(B)}\right]^{-1}
		\\
		&\times
		\left[\sum_{k,l=1}^6 Q(a_k,b_l)|a_k\rangle\langle a_k|\otimes|b_l\rangle\langle b_l|\right]
		\\
		&\times
		\left[U^{(A)}L^{(A)}\otimes U^{(B)}L^{(B)}\right]^{-\dag}.
	\end{aligned}
	\end{equation}
	In this formula, we use normalized states and a correspondingly renormalized distribution, given as follows:
	\begin{equation}
	\begin{aligned}
		|\tilde a_k\rangle\langle \tilde a_k|
		=&\frac{L^{(A)-1}U^{(A)\dag}|a_k\rangle\langle a_k|U^{(A)}L^{(A)-\dag}}{\langle a_k|U^{(A)}L^{(A)-\dag}L^{(A)-1}U^{(A)\dag}| a_k\rangle},
		\\
		|\tilde b_l\rangle\langle \tilde b_l|
		=&\frac{L^{(B)-1}U^{(B)\dag}| b_l\rangle\langle b_l|U^{(B)}L^{(B)-\dag}}{\langle b_l|U^{(B)}L^{(B)-\dag}L^{(B)-1}U^{(B)\dag}| b_l\rangle},
		\quad\text{and}
		\\
		Q(\tilde a_k,\tilde b_l)=&Q(a_k,b_l)
		\langle a_k|U^{(A)}L^{(A)-\dag}L^{(A)-1}U^{(A)\dag}| a_k\rangle
		\\&\times
		\langle b_l|U^{(B)}L^{(B)-\dag}L^{(B)-1}U^{(B)\dag}| b_l\rangle.
	\end{aligned}
	\end{equation}
	Similarly to the previous two-step description, we depict the resulting transformation of both local states from standard form over rotations $[U^{(A)}\otimes U^{(B)}]^{-1}$ to the final boost transformations $[L^{(A)}\otimes L^{(B)}]^{-1}$ in Fig. \ref{fig:LocalStateTransform}.

	We can also confirm that the POVM element expressed by this quasidistribution and the corresponding local states does, within the numerical precision, exactly describe the previously reconstructed POVM element in Fig. \ref{fig:BasisReconstrct}.
	This demonstrates the successful representation of an entangled POVM via quasidistributions.

\begin{figure*}
	\includegraphics[width=.36\textwidth]{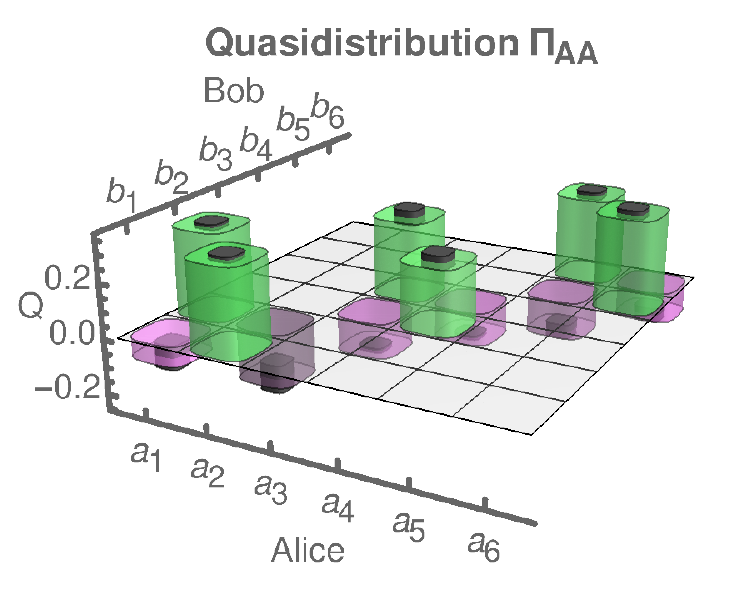}
	\hfill
	\includegraphics[width=.22\textwidth]{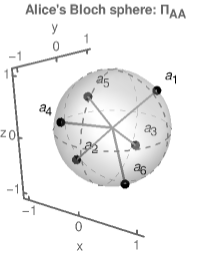}
	\hfill
	\includegraphics[width=.22\textwidth]{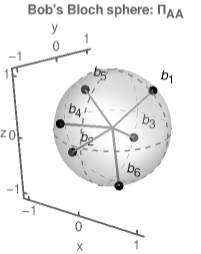}
	\\\vspace{2ex}
	\includegraphics[width=.36\textwidth]{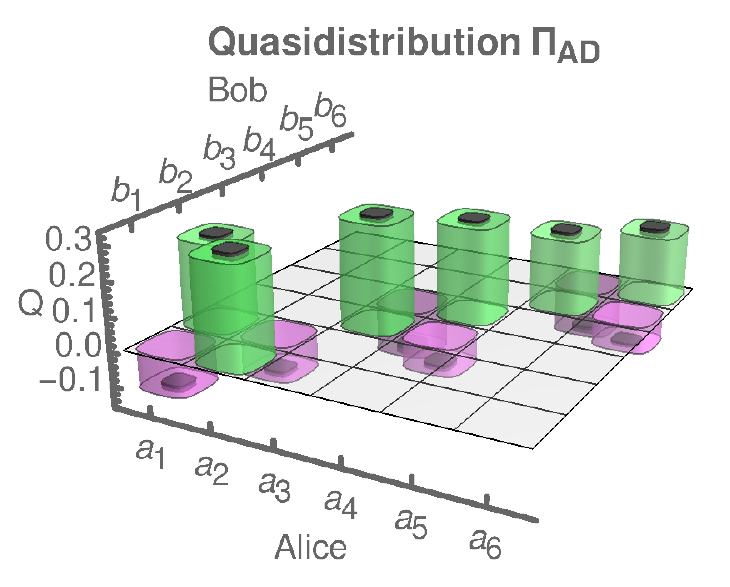}
	\hfill
	\includegraphics[width=.22\textwidth]{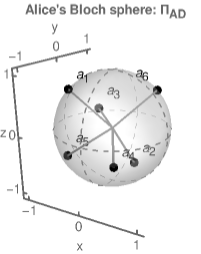}
	\hfill
	\includegraphics[width=.22\textwidth]{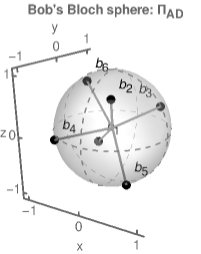}
	\\\vspace{2ex}
	\includegraphics[width=.36\textwidth]{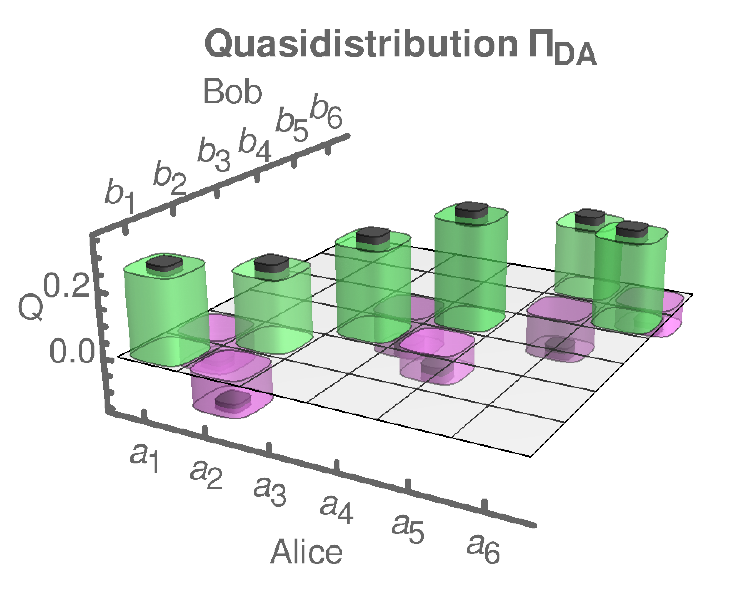}
	\hfill
	\includegraphics[width=.22\textwidth]{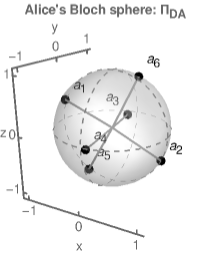}
	\hfill
	\includegraphics[width=.22\textwidth]{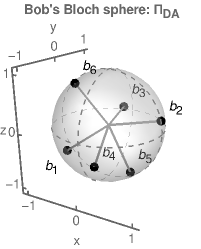}
	\\\vspace{2ex}
	\includegraphics[width=.36\textwidth]{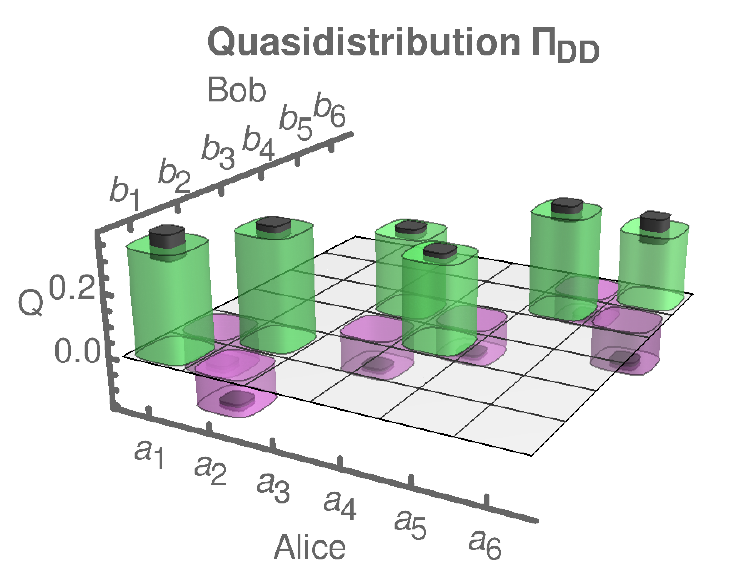}
	\hfill
	\includegraphics[width=.22\textwidth]{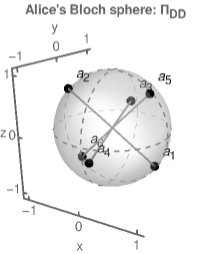}
	\hfill
	\includegraphics[width=.22\textwidth]{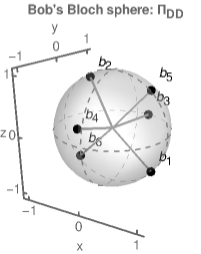}
	\caption{%
		Reconstructed quasidistributions (left column) for all POVM elements, including a one-standard-deviation error margin (black bars).
		The right column shows the corresponding local states for the decomposition according to Eq. \eqref{eq:LocDecomp}.
		In comparison with the ideal cases (Fig. \ref{fig:BellIdealquasi}), one can observe here the same general structure of POVM elements.
		In theory, the maximal negativity is $-1/6\approx-0.17$.
		Here, we find the highest negativities as $-0.20\pm0.06$ (mind the error margin), $-0.14\pm0.01$, $-0.13\pm0.02$, and $-0.14\pm0.02$ for $AA$, $AD$, $DA$, and $DD$, respectively.
	}\label{fig:Results}
\end{figure*}

\subsection{Error propagation}
\label{subsec:Uncertainties}

	The methods described so far have been applied to estimate mean values.
	To determine uncertainties, a random sample for a Monte Carlo error propagation is prepared.
	Each sample element undergoes the aforementioned processing, allowing one to estimate the resulting fluctuations.
	The ensuing error estimates for our quasidistributions are depicted in Fig. \ref{fig:Results}.

	To implement the error propagation, a sample of $10\,000$ relative coincidence matrices $[P_k(a,b)]_{k\in\{AA,AD,DA,DD\}}$ is generated for each probe-state setting $(a,b)$.
	This sample is distributed with a mean that corresponds to previously determined relative frequencies $\mu_k=P_k(a,b)=E_k(a,b)/E(a,b)$ [Eq. \eqref{eq:RelFreq}].
	Fluctuations are implemented through the covariance matrix of the counting statistics $\Sigma_{k,k'}=[\delta_{k,k'}P_k(a,b)-P_k(a,b)P_{k'}(a,b)]/[E(a,b)-1]$ to describe the standard deviation as well as cross-correlations in the data.
	These uncertainties are multiplied by $1.05$ to provide an extra $5\%$ error margin as a safeguard to counter common issues, such as undersampling.
	The sample elements generated in this manner are further chosen to be normalized and nonnegative as they resemble probabilities $P_k(a,b)\geq0$ and $\sum_{k}P_k(a,b)=1$.

	As mentioned before, each sample element is treated with the reconstruction approaches established in Secs. \ref{sec:DataProcessingI} and \ref{sec:DataProcessingII}.
	The standard deviation, for example, of the resulting sample of quasidistributions then provides the error margin, as depicted in Fig. \ref{fig:Results}.
	This concludes the full reconstruction from detector-tomography raw data to quasidistributions.


\section{Discussion and conclusion}
\label{sec:Conclusion}

	After comprehensively presenting the data processing approach, our conclusions from this reconstruction are presented in this section, together with the implications pertaining to the entangled nature of the realized BSM.
	A brief discussion is provided in Sec. \ref{subsec:Results}.
	An additional comparison with separable POVMs is done in Sec. \ref{subsec:SepCompare}.
    Moreover, a probe-state method to certify POVM entanglement is introduced and applied to high-dimensional and multipartite systems, additionally allowing us to probe noise robustness of detector entanglement when scaling the system size.
	Eventually, we summarize the findings of the paper in Sec. \ref{subsec:Summary}.

\subsection{Results}
\label{subsec:Results}

	The entangled POVMs in Fig. \ref{fig:Results} are locally described via Eq. \eqref{eq:LocDecomp}, however, requiring the depicted negativities in the joint distribution to capture the detector entanglement of the experimentally implemented device.
	Furthermore, these results structurally relate quite well to the ideal POVM elements that one expects for unperturbed BSMs (Fig. \ref{fig:BellIdealquasi}).
	For instance, the nonlocal (negative) and local (positive) contributions are found in the same pattern that one can see in the theory plots.

	In terms of statistical significance, we find that the most significant negativities are $15$ standard deviations for $\Pi_{AA}$, $17$ standard deviations for $\Pi_{AD}$, $8$ standard deviations for $\Pi_{DA}$, and $16$ standard deviations for $\Pi_{DD}$ below the classical threshold of zero.
	Therefore, our results show a highly significant POVM entanglement of the implemented detection scheme.
	(Note that those highest significances do not necessarily coincide with the ones that exhibit the highest absolute negativity reported in Fig. \ref{fig:Results}.)
	Furthermore, using the same POVM element order, the cumulative negativities, i.e., the sums over all negative entries of $Q$, are $-0.77\pm0.08$, $-0.65\pm0.03$, $-0.65\pm0.05$, and $-0.72\pm0.03$.
	For comparison, the perfect case yields $-1$ for all Bell-type POVM elements (Fig. \ref{fig:BellIdealquasi} with six negative contributions with the value $-1/6$), not being drastically larger than what we find for our data.

	We emphasize that, except for accounting for unphysical eigenvalues of POVM elements by mixing with separable noise, no corrections for imperfections, such as deconvolutions of attenuations and postprocessing for other sources of noise, have been carried out.
	Still, a highly significant verification of detector entanglement that is essential for quantum information processing was confirmed with our methodology.
	Furthermore, the framework provides an intuitive (visual) signature of entanglement of detectors and yields a unified foundation with entanglement of states by virtue of analogous entanglement quasiprobability methods \cite{SMBBS19}.
	Also, since such analog methods for states extend to multipartite and qudit entanglement \cite{SW18}, the POVM framework discussed here is similarly extendable to high-dimensional scenarios.

\begin{figure}
	\includegraphics[width=.49\columnwidth]{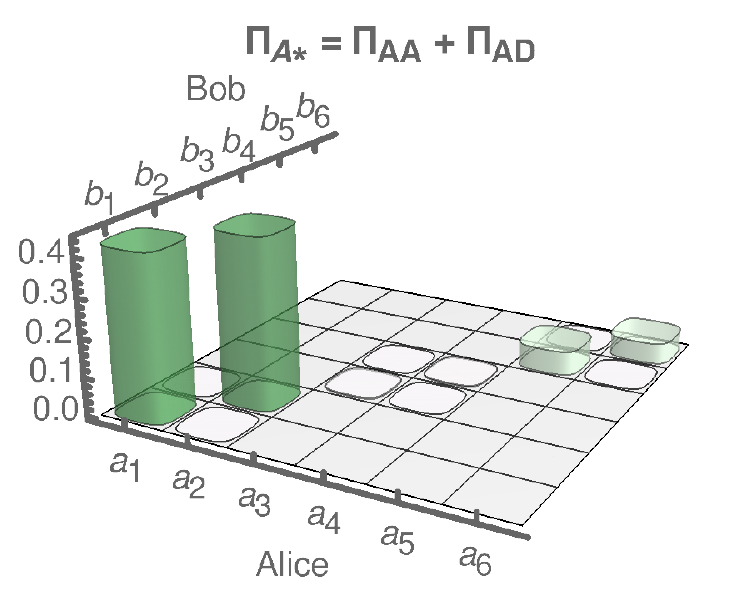}
	\includegraphics[width=.49\columnwidth]{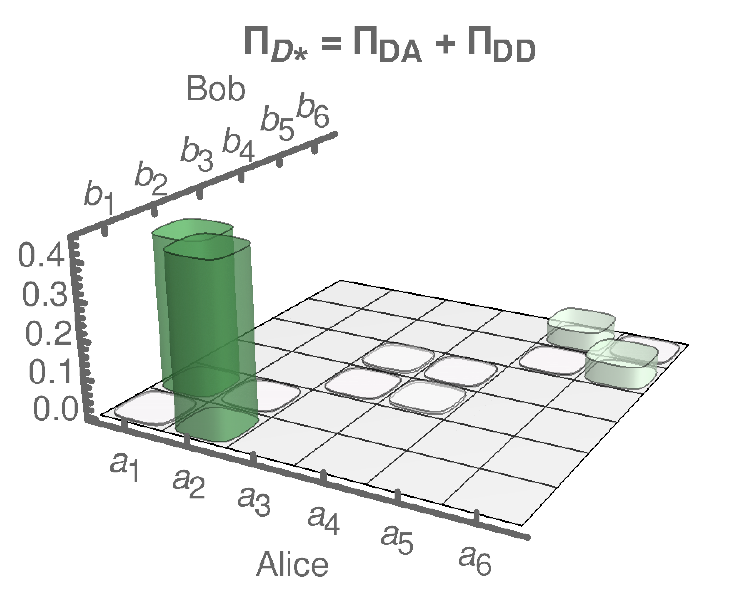}
	\\\vspace{2ex}
	\includegraphics[width=.49\columnwidth]{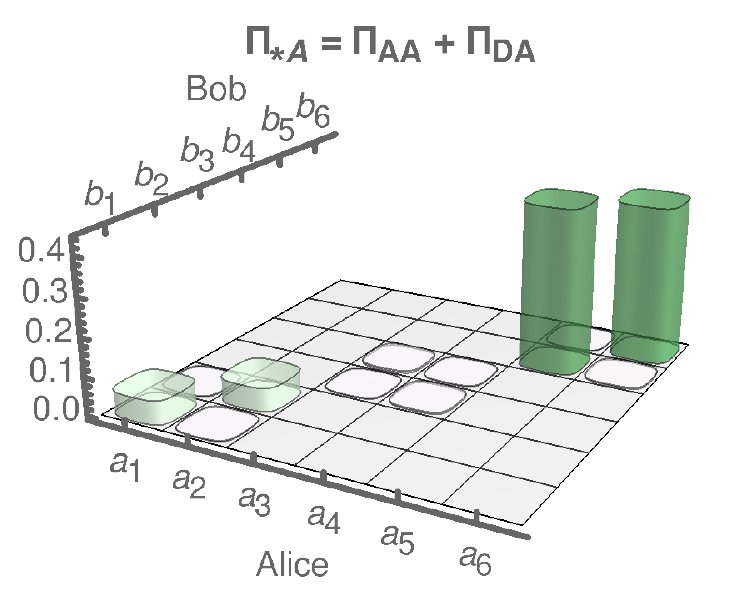}
	\includegraphics[width=.49\columnwidth]{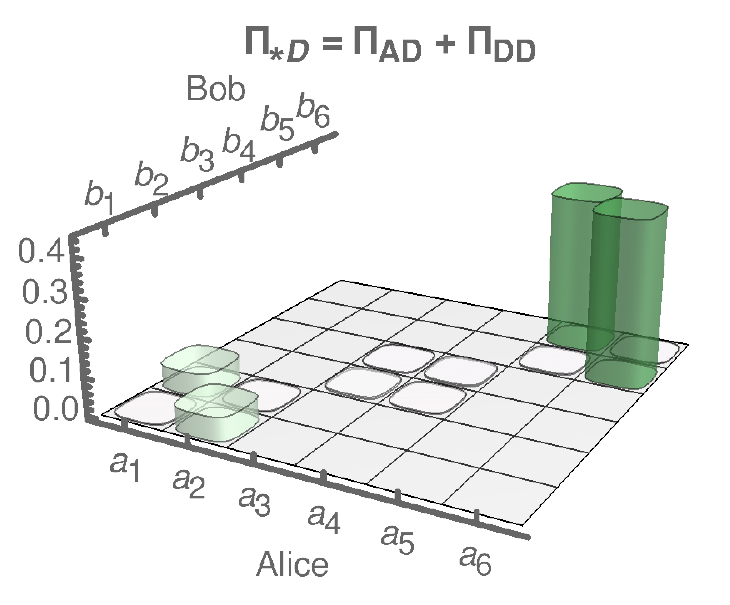}
	\caption{%
		Quasidistributions for POVMs that combine two Bell-like states and are therefore expected to be separable.
		Our reconstruction correctly reveals this feature.
	}\label{fig:SepCombinations}
\end{figure}

\subsection{Comparison with separable POVMs}
\label{subsec:SepCompare}

	For completeness, we may also show that separable POVMs truly lead to nonnegative---i.e., classical---distributions.
	To this end, it is worth mentioning the known fact that the uniform mixture of a Bell state with any of the other Bell states results in a separable operator.
	Thus, we might mix our data accordingly to probe if this indeed produces non-entangled POVMs.

	For example, we can combine POVM elements which are identical in one of the indices to produce new POVMs, such as $\{\Pi_{A \ast}=\Pi_{AA}+\Pi_{AD},\Pi_{D \ast}=\Pi_{DA}+\Pi_{DD}\}$ and $\{\Pi_{\ast A}=\Pi_{AA}+\Pi_{DA},\Pi_{\ast D}=\Pi_{AD}+\Pi_{DD}\}$.
	In terms of data, this means adding counts $E_k(a,b)$ accordingly and following the same reconstruction approach as carried out for the BSM.
	The results of this treatment can be found in Fig. \ref{fig:SepCombinations}.
	Indeed, the quasidistributions for both sets of POVMs appear to be nonnegative, as one expects for separable POVMs.
	In terms of our device, this means that it is vital to record the outcomes of both detectors to be able to harness the detector entanglement in quantum protocols as the loss of that information results in a local model of the measurement device.

\subsection{Entanglement-probing states for POVM entanglement}
\label{subsec:witnessing}

	We consider further examples to highlight the versatility of our approach.
	Before that, we make the observation that the entanglement of a POVM element $\Pi$ can be simply related to entanglement of states.
	That is, the rescaled operator
	\begin{equation}
		\rho_\text{POVM}=\frac{\Pi}{\mathrm{tr}(\Pi)}
	\end{equation}
	describes a proper quantum state.
	For infinite-dimensional systems, such a normalization may not be possible;
	still, in such cases, finite-dimensional subspaces that show the entanglement may be constructed \cite{SV09finite}.

	For entangled states, here $\rho_\text{POVM}$, one can construct entanglement witnesses to probe.
	Specifically, an operator $L$ exist such that
	\begin{equation}
		\mathrm{tr}(\rho_\text{POVM} L)>g_{\max}(L),
	\end{equation}
	where $g_{\max}(L)$ is the maximal expectation value of $L$ for separable states;
	see Ref. \cite{SV13} and references therein.
	Furthermore, it was shown that $L$ can be translated with the identity and rescaled with a positive factor without altering the relation in the above inequality \cite{SV13}, e.g.,
	\begin{equation}
		\label{eq:witnessing}
		\frac{\mathrm{tr}(\Pi \rho_\text{probe})}{\mathrm{tr}(\Pi)}>g_{\max}(\rho_\text{probe}),
	\end{equation}
	where $\rho_\text{probe}=\alpha\mathbbm1+\beta L$ [hence $g_{\max}(\rho_\text{probe})=\alpha+\beta g_{\max}(L)$] is chosen such that it represents a valid probe state for the POVM element $\Pi$.

	Now, $g_{\max}(\rho_\text{probe})$ can now be interpreted as the maximal outcome that a separable measurement can produce for the probe state.
	Therefore, the inequality \eqref{eq:witnessing} shows that the probe state produces an outcome for the POVM element $\Pi$ under study that exceeds the expectation of separable POVMs.

	The probe-state method allows us to study the scaling behavior of entanglement as a function of the local dimension as well as the number of parties in a multipartite setting.

	We begin with a multipartite setting in which separability is based on projectors of the form $\bigotimes_{j=1}^n|a^{(j)}\rangle\langle a^{(j)}|$ for $n$ parties.
	As a particular example, we consider a Greenberger--Horne--Zeilinger (GHZ) projector mixed with white noise (modeled through the identity)
	\begin{equation}
		\Pi=\varepsilon\mathbbm{1}+(1-\varepsilon)|\text{GHZ}\rangle\langle\text{GHZ}|,
	\end{equation}
	with $\mathrm{tr}(\Pi)=\varepsilon 2^n+(1-\varepsilon)$ and $|\text{GHZ}\rangle=(|0\rangle^{\otimes n}+|1\rangle^{\otimes n})/\sqrt2$.
	Thus, we have an ideal GHZ measurement for $\varepsilon=0$, and $\varepsilon\to1$ yields the identity that is non-entangled.
	As the entanglement-probing state, we consider
	\begin{equation}
		\rho_\mathrm{probe}=\frac{\mathbbm1+(|0\rangle\langle1|)^{\otimes n}+(|1\rangle\langle0|)^{\otimes n}}{2^n}.
	\end{equation}
	Using the exact results from the Appendix, we obtain
	\begin{equation}
		g_{\max}(\rho_\text{probe})=\frac{1+2^{1-n}}{2^n}.
	\end{equation}
	The left-hand side of the inequality \eqref{eq:witnessing} reads
	\begin{equation}
		\frac{\mathrm{tr}(\Pi \rho_\text{probe})}{\mathrm{tr}(\Pi)}=\frac{\varepsilon 2^n+2(1-\varepsilon)}{2^n[\varepsilon 2^n+(1-\varepsilon)]}.
	\end{equation}

	Therefore, inequality \eqref{eq:witnessing} holds true when the noise is upper bounded as
	\begin{equation}
		\varepsilon<\frac{2^{n-1}-1}{3\times 2^{n-1}-1}.
	\end{equation}
	In the limit of a macroscopic number of qubits $n\to\infty$, this means that up to $33.3\%$ white noise can be tolerated in the detection without losing the entanglement properties of the POVM element under study.
	By comparison, we have $20\%$ noise resilience for the bipartite case, $n=2$.

	Next, we explore a two-qudit system with a local dimension $d$ for each.
	Analogously to the previous example, the POVM element based on maximally entangled (ME) projectors is considered,
	\begin{equation}
		\Pi=\varepsilon\mathbbm1+(1-\varepsilon)|\text{ME}\rangle\langle\text{ME}|,
	\end{equation}
	with $|\text{ME}\rangle=d^{-1/2}\sum_{k=0}^{d-1}|k\rangle\otimes|k\rangle$.
	Here, the probe state is chosen as
	\begin{equation}
		\rho_\text{probe}=\frac{1}{d^2}\left(
			\mathbbm 1+\sum_{\substack{k,l=0 \\ k\neq l}}^{d-1}|k\rangle\langle l|\otimes|k\rangle\langle l|
		\right).
	\end{equation}
	Again, the general results in the Appendix provide the bound
	\begin{equation}
		g_{\max}(\rho_\text{probe})=\frac{1+\left(1-\frac{1}{d}\right)}{d^2}.
	\end{equation}
	Similarly to the previous example, the inequality \eqref{eq:witnessing} renders it possible to find the limit of the noise contribution to ensure entanglement of the POVM element,
	\begin{equation}
		\varepsilon<\frac{d-2+\frac{1}{d}}{d^2-2+\frac{1}{d}}.
	\end{equation}
	For two-qubit measurements $d=2$, we have a measurement-noise threshold of $20\%$, and the noise sensitivity of the POVM element increases as $\sim1/d$ for local dimensions $d\to\infty$.

\subsection{Summary}
\label{subsec:Summary}

    In summary, we introduced a framework to assess entanglement of POVMs.
    Based on our definition of a separable POVM in terms of nonnegative mixtures of local (i.e., product) projectors, entangled POVMs cannot exhibit such a local representation.
    Rather, nonlocal coherence contributes to the recorded measurement outcome from entangled POVMs, even when separable states are measured.
    As an intuitive approach, it was shown that these nonlocal measurement features can be represented in terms of pseudomixtures, meaning a local-like representation (preserving product projectors) is possible when allowing for mixing ratios that include negative contributions, while being strictly nonnegative for classically correlated detection schemes.
    The negativity of the thereby defined joint quasidistribution constitutes a necessary and sufficient criterion for POVM entanglement.
    Complementing quasidistributions, a probe-state method was devised and applied theoretically to qudits and multipartite detection scenarios that leads to measurement outcomes that exceed the capabilities of separable detectors.

    We studied in detail BSMs because of their fundamental and applied importance.
    We provided the comprehensive step-by-step reconstruction from raw detector-tomography data to fully reconstructed quasidistributions, including error estimates.
    With high statistical significance, it was then certified that the experimental detection scheme under study does include negativities in the quasidistributions of each POVM element.
    This means that the operation of this detector cannot be explained in terms of local coherence effects alone for any of the possible measurement outcomes, identifying entanglement as a essential quantum resource for the function of this device.
    As a counterexample, we also showed that our data processing method correctly leads to nonnegative joint distributions for separable POVMs, thus admitting a local measurement model.

    Therefore, our highly sensitive and comparably easily accessible diagnostic tool offers an alternative approach to characterizing quantum detectors regarding their quantum-correlation properties.
    This not only renders it possible to decide the fundamental question whether a local description of a specific quantum measurement is possible but also provides a practical means to assess the nonlocal performance of detection devices for quantum-technological applications.


\begin{acknowledgments}
    The authors acknowledge the kind availability of E. Roccia, M. Sbroscia, and M. G. Genoni in sharing the data for our analysis.
	This work was supported by the FET-OPEN-RIA project STORMYTUNE (Grant Agreement No. 899587).
	J.S. acknowledges financial support from the Deutsche Forschungsgemeinschaft (DFG, German Research Foundation) through the Collaborative Research Center TRR~142 (Project No. 231447078, project C10).
	E.A. acknowledges funding from the Der Wissenschaftsfonds FWF (Fonds zur F\"{o}rderung der wissenschaftlichen Forschung) Lise Meitner-Programm (M3151).
\end{acknowledgments}

\appendix*
\section{Exact upper bounds for test operators}
\label{app:ExactSolutions}

	Consider an operator
	\begin{equation}
		\Lambda=\sum_{k,l=0}^{d-1}\left(|k\rangle\langle l|\right)^{\otimes n}-\sum_{k=0}^{d-1}\left(|k\rangle\langle k|\right)^{\otimes n}
	\end{equation}
	acting on $n$, $d$-dimensional parties.
	Our goal is to find the maximal overlap of this operator with pure product states.
	To this end, the separability eigenvalue equation from Ref. \cite{SV13} is applied.
	For the $j$th subsystem and the $k$th component thereof, the equation reads
	\begin{equation}
    \label{eq:A2}
	\begin{aligned}
		&\prod_{\substack{i=1 \\ i\neq j}}^na_k^{(i)\ast}
		\left(
			\sum_{l=0}^{d-1}a_l^{(1)}\cdots a_l^{(n)}
			-a_k^{(1)}\cdots a_k^{(n)}
		\right)
		\\
		=&\langle a^{(1)},\ldots,a^{(j-1)},k,a^{(j+1)},\ldots,a^{(n}|\Lambda|a^{(1)},\ldots,a^{(n)}\rangle
		\\
		=&g\langle k|a^{(j)}\rangle=ga^{(j)}_k,
	\end{aligned}
	\end{equation}
	where $|a^{(1)},\ldots,a^{(n)}\rangle$ is an $n$-partite tensor product of normalized vectors and $g$ denotes the sought-after separability eigenvalue that yields the maximum.

	Multiplying the Eq. \eqref{eq:A2} by $a_k^{(j)\ast}$ results in an expression on the leftmost side of the equation that is independent of $j$ and $g|a_k^{(j)}|^2$ on the rightmost side.
	Thus, $|a_k^{(j)}|=r_k$ is constant with respect to $j$.
    Note that the normalization now reads $\nu=\sum_{k=0}^{d-1}r_k^2=1$.
	Further, using the summed phase $\Phi_k=\sum_{j=1}^n\arg(a_{k}^{(j)})$ as an abbreviation, the separability eigenvalue reads
	\begin{equation}
	\begin{aligned}
		g
		=&\langle a^{(1)},\ldots,a^{(n)}|\Lambda|a^{(1)},\ldots,a^{(n)}\rangle
		\\
		=&\left|\sum_{k=0}^{d-1}r_k^n e^{i\Phi_k}\right|^2-\sum_{k=0}^{d-1}r_k^{2n}.
	\end{aligned}
	\end{equation}
	This is maximal when $e^{i\Phi_k}$ is constant with respect to $k$, summing only terms that are in phase.

	Finally, we can optimize $g$ under the constraint of normalization $\nu$ using the method of Lagrangian multipliers,
	\begin{equation}
		0=\frac{\partial g}{\partial r_l}-\mu\frac{\partial \nu}{\partial r_l}.
	\end{equation}
	After some straightforward algebra, the resulting identity can be recast into the form
	\begin{equation}
		\sum_{k=0}^{d-1}r_{k}^n=\frac{\mu}{n}r_l^{2-n}+r_l^n.
	\end{equation}
	Another possible solution is $r_l=0$.
	Again, the left-hand side of the identity is independent of $l$, implying a constant $r_l=1/\sqrt{d'}$ to satisfy the normalization when $d'$ many $r_k$s are non-zero.
	Inserting this result, we obtain
	\begin{equation}
		g=\frac{d'(d'-1)}{d^{\prime n}}.
	\end{equation}
	For $n=2$, $g$ is strictly monotonic increasing with $d'$, and $g_{\max}$ is obtained for $d'=d$.
	The contrary is true for $n>2$ and $d'>1$ ($d'=1\Rightarrow g=0$), and $g_{\max}$ is thus obtained for $d'=2$.

	As a final remark, it is worth mentioning that the operator $\Lambda$ has a maximal (ordinary) eigenvalue $d-1$ and a minimal eigenvalue $-1$.

\bibliographystyle{refstyle}
\bibliography{refs}

\end{document}